\RequirePackage{fix-cm}
\documentclass[smallcondensed]{svjour3}                     
\smartqed  
\usepackage{graphicx}
\usepackage{newtxtext,newtxmath} 
\usepackage[utf8]{inputenc}
\usepackage{subfig}
\usepackage{siunitx}
\usepackage{amsmath,amsfonts,xcolor,stmaryrd}
\usepackage{nicefrac}
\usepackage{hyperref}
\hypersetup{
	colorlinks=true,
	linkcolor=blue,
	citecolor=blue,
	filecolor=magenta,      
	urlcolor=cyan,
}

\usepackage{eqnarray}

\usepackage[font=small]{caption}
\usepackage{array,contour}
\contourlength{0.02em}

\graphicspath{{./}}
\newcommand{\coma}{\>\text{,}}
\newcommand{\point}{\>\text{.}}
\newcommand{\nv}{\text}
\newcommand{\ve}[1]{\boldsymbol{#1}} 
\newcommand{\T}{^\top}
\newcommand{\dv}{\,\text{d}V}
\newcommand{\da}{\,\text{d}A}

\newcommand{\elli}{{\ell_\text{i}}}
\newcommand{\ellc}{{\ell_\text{c}}}
\newcommand{\gc}{\mathcal{G}_\text{c}}
\newcommand{\gcb}{\mathcal{G}_\text{c}^\text{b}}
\newcommand{\gci}{\mathcal{G}_\text{c}^\text{i}}

\newcommand{\gcih}{\hat{\mathcal{G}}_\text{c}^\text{i}}

\newcommand{\gammai}{\Gamma^\text{i}}
\newcommand{\gammac}{\Gamma^\text{c}}
\newcommand{\gammalc}{\Gamma^{\ell_\text{c}}}
\newcommand{\gammali}{\Gamma^{\ell_\text{i}}}

\newcommand{\rgauss}{\mathcal{G}_\text{c}}

\newcommand{\eto}{\tilde{E}_1}
\newcommand{\eo}{E_1}
\newcommand{\ett}{\tilde{E}_2}
\newcommand{\et}{E_2}
\newcommand{\nuto}{\tilde{\nu}_1}
\newcommand{\nuo}{\nu_1}
\newcommand{\nutt}{\tilde{\nu}_2}
\newcommand{\nut}{\nu_2}

\DeclareMathOperator{\sign}{sign}
\usepackage{scalerel}

\newcommand{\sig}{\boldsymbol{\sigma}}

\newcommand{\eps}{\boldsymbol{\varepsilon}}
\newcommand{\jumpeps}{\llbracket\eps\rrbracket}

\newcommand{\etaf}{\eta_\text{f}}
\newcommand{\jump}[1]{\llbracket #1 \rrbracket}
\renewcommand{\bold}[1]{\boldsymbol{#1}}
\newcommand{\nob}{\ve{n}^\text{b}}
\newcommand{\noi}{\ve{n}^\text{i}}
\newcommand{\R}{\mathbb{R}}

\newcommand{\opsi}[3]{\overset{#1}{\psi}\mathstrut^\text{#2}_{#3}}
\newcommand{\osig}[2]{\overset{#1}{\sig}\mathstrut^{#2}}

\renewcommand{\Omega}{\varOmega}
\renewcommand{\Gamma}{\varGamma}
\renewcommand{\Sigma}{\varSigma}
\renewcommand{\Psi}{\varPsi}
\renewcommand{\Phi}{\varPhi}
\renewcommand{\Pi}{\varPi}

\newcommand{\legline}[1]{\raisebox{-0.1cm}{\protect\includegraphics{line#1.pdf}}}

\usepackage{multirow}
\usepackage{booktabs}
\usepackage{blindtext}

\journalname{Archive of Applied Mechanics}
\begin{document}

\title{Phase-field modeling of fracture in
	heterogeneous materials: jump conditions, convergence and crack propagation
}

\titlerunning{Phase-field modeling of fracture in heterogeneous materials}        

\author{Arne Claus Hansen-Dörr        \and
		Jörg Brummund 			  	  \and
        Markus Kästner 
}


\institute{A.\,C. Hansen-Dörr 
           \and
           J. Brummund\and
           M. Kästner \at
           Institute of Solid Mechanics, TU Dresden, 01062 Dresden, Germany \\
           \email{markus.kaestner@tu-dresden.de}   
}

\date{Received: 3 April 2020 / Accepted: 19 August 2020}

\maketitle

\begin{abstract}
In this contribution, a variational diffuse modeling framework for cracks in heterogeneous media is presented. A static order parameter smoothly bridges the discontinuity at material interfaces, while an evolving phase-field captures the regularized crack. The key novelty is the combination of a strain energy split with a partial rank-I relaxation in the vicinity of the diffuse interface. The former is necessary to account for physically meaningful crack kinematics like crack closure, the latter ensures the mechanical jump conditions throughout the diffuse region. 

The model is verified by a convergence study, where a circular bi-material disc with and without a crack is subjected to radial loads. For the uncracked case, analytical solutions are taken as reference. In a second step, the model is applied to crack propagation, where a meaningful influence on crack branching is observed, that underlines the necessity of a reasonable homogenization scheme. The presented model is particularly relevant for the combination of any variational strain energy split in the fracture phase-field model with a diffuse modeling approach for material heterogeneities.  

\end{abstract}

\section{Introduction}
\label{intro}

Modern engineering simulation challenges comprise the prediction of failure, which is one of the most severe mechanism affecting the bearing capacity. The phase-field method for the simulation of crack growth proved to be a powerful tool because it incorporates crack nucleation and arrest, as well as branching and merging of cracks~\cite{bourdin_variational_2008,miehe_phase_2010,kuhn_continuum_2010}. Based on the variational approach to brittle fracture~\cite{francfort_revisiting_1998}, it regularizes the underlying energy functional~\cite{bourdin_numerical_2000} and approximates the crack by an auxiliary scalar field, which is referred to as crack phase-field. The phase-field smoothly bridges the intact and fully broken state by introducing a length scale $\ellc$. The approach is consistent with the energetic cracking criterion introduced by Griffith~\cite{griffith_phenomena_1921}.
The phase-field method for crack modeling allows for fixed meshes where the element edges do not have to be aligned with the crack path, i.e. remeshing is avoided when the crack changes its direction or branches. 

The approach for a non-conforming description can be extended to a more general setting, where all discontinuities within a heterogeneous structure are captured in such a way. Cumbersome preprocessing such as manual identification of heterogeneities and meshing can be skipped or at least simplified if the structure is for example obtained from direct imaging~\cite{nguyen_phase_2015,nguyen_phase-field_2016,nguyen_phase-field_2017}. Prominent representatives are the extended finite element method~\cite{sukumar_modeling_2001,belytschko_review_2009,fries_extendedgeneralized_2010} or the finite cell method~\cite{parvizian_finite_2007,duster_finite_2008}. The latter is among the category of fictitious domain methods, where the geometry is fully embedded within a larger domain~\cite{hansbo_unfitted_2002,schillinger_unfitted_2011,hennig_diffuse_2019}.

Another way to incorporate material heterogeneities is closely related to the phase-field model for cracks. A (multi)phase-field, which introduces a diffuse region $\gammali$, with a characterizing width length scale $\elli$, along the formerly sharp interface $\gammai$, marks the individual subdomains, and offers a broad range of applications in the field of thermodynamics, chemistry and mechanics~\cite{steinbach_multi_2006,schneider_small_2015,schneider_phase-field_2016}. The individual phase-field parameters are often referred to as order parameters. Far away from the interface, the properties of the constituents are recovered. Within the diffuse region, additional assumptions such as the treatment of strong discontinuities of the stress or strain are required, which will be highlighted in the following. 

In classical homogenization theory, the Voigt and Reuss limits play a major role. Transferring the corresponding assumptions of a constant strain or a constant stress within multiple constituents to the diffuse interface region implies, that either the strain or stress does not exhibit a jump and the corresponding quantities are equal for each phase. It can be shown, that these assumptions converge towards the sharp interface limit~\cite{steinbach_multi_2006}. However, better approaches exist, which account for the stress and strain discontinuity across the interface simultaneously~\cite{hennig_diffuse_2019,mosler_novel_2014,schneider_phase-field_2015,herrmann_multiphase-field_2018}. The improvement manifests itself in better convergence rates and the fact, that the kinematics of an energy driven, moving interface are captured correctly~\cite{kiefer_numerical_2017}. The improved homogenization scheme is often referred to as partial rank-I relaxation, which will become apparent in the remainder of the paper, and yields a pointwise fulfillment of the equilibrium and compatibility within the diffuse region.

We presented a modeling framework~\cite{hansen-dorr_phase-field_2020}, where a crack phase-field and a static order parameter were used to simulate crack branching and deflection in heterogeneous media. Besides a qualitatively and quantitatively good agreement with analytical predictions from linear elastic fracture mechanics~\cite{he_crack_1989}, we observed inconsistencies of the modeling results when comparing the diffuse interface approach to a sharp interface simulation. We presumed, that the Voigt-Taylor homogenization approach, which was incorporated in the diffuse region, biased the crack phase-field driving force in a way, that a straight crack instead of a deflection at the interface was favored. As discussed above, the homogenization scheme significantly influences the energetic driving force of the phase transition. 

This work extends our diffuse modeling framework~\cite{hansen-dorr_phase-field_2020} in a way, that a partial rank-I relaxation is incorporated within the diffuse interface region. The key novelty of the present paper is the variationally consistent combination of the aforementioned homogenization approach with an additive strain energy decomposition, which is necessary to account for physically meaningful crack behavior such as the tension compression asymmetry. It will be shown, that the novel scheme is superior to the classical Voigt-Taylor assumption. 

The paper is structured as follows. Section~\ref{sec:theory} provides a variational formulation of the modeling approach. Subsequently in Section~\ref{sec:numex}, a convergence study and crack propagation simulations provide an insight into the functionality and advantages of the model. Final conclusions summarize the paper in Section~\ref{sec:conclusion}. Additional information concerning technical details are provided in Appendices~\ref{app:stress} to \ref{app:traction}. 

\section{Phase-field model for cracks and diffuse heterogeneities}
\label{sec:theory}
\begin{figure}[t]
	\centering
	\subfloat[Discrete crack representation]{\label{fig:disccrack}\includegraphics{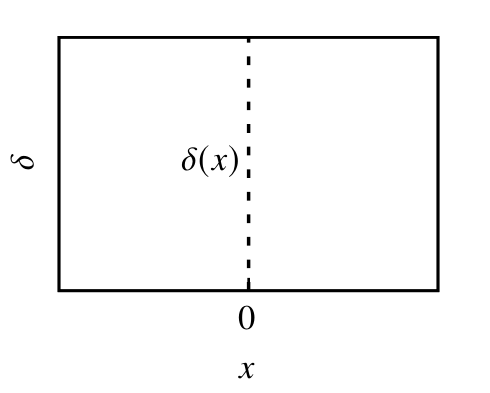}}
	\subfloat[Regularized crack representation]{\label{fig:regcrack}\includegraphics{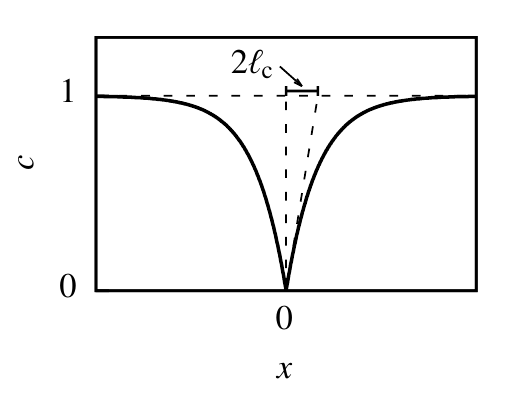}}
	\caption{In (a), the location of the sharp crack is described by the \textsc{Dirac} distribution. The regularized representation using an exponential function is depicted in (b). The length scale parameter $\ellc$ controls the width of the transition region from $c=0$ to $c=1$.
	}
	\label{fig:discreg}
\end{figure}
  
\subsection{Phase-field modeling of cracks} 
\begin{figure*}[t]
	\centering
	\subfloat[Heterogeneous material with a phase-field crack $\gammalc$ and a diffuse interface $\gammali$]{
		\label{fig:discif}
		\includegraphics{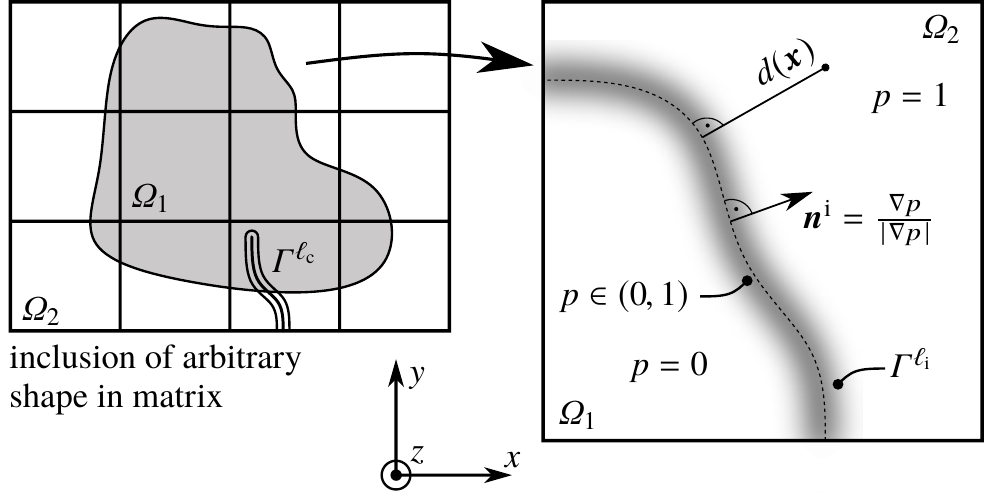}
	}
	\subfloat[Order parameter $p$]{
		\label{fig:orderparam}
		\begin{minipage}[b][5.3cm][t]{0.33\textwidth}
			\centering
			\footnotesize
			\includegraphics{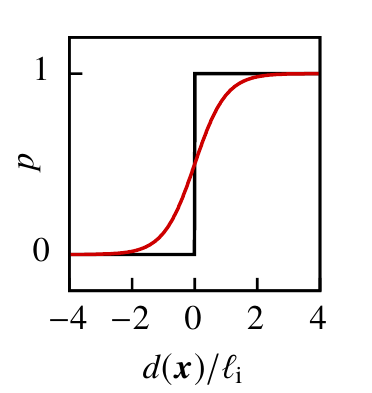}
			\begin{tabular}{l}
				\legline{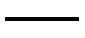} sharp interface \\ 
				\legline{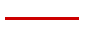} $\tanh$ transition
			\end{tabular}
	\end{minipage}}
	\caption{The diffuse modeling approach is exemplarily depicted in (a), where a heterogeneity is embedded in a regular mesh. The dashed interface mid-surface is described by the level-set $d(\ve{x})\equiv0$ of the signed distance function. The sharp interface $\gammai$ is smoothed by a hyperbolic tangent transition (b), and becomes a diffuse region $\gammali$.
	}
	\label{fig:discdiff}
\end{figure*}

The investigations presented in this contribution build on a fully diffuse framework for crack propagation simulations in heterogeneous materials. The crack is described following the phase-field approach. Suppose, a one-dimensional rod $x\in(-\infty,\infty)$ of infinite length is cracked at $x=0$. The sharp crack location can be fixed using the \textsc{Dirac} distribution, cf. Figure~\ref{fig:disccrack}. In contrast to a sharp crack, the phase-field approach introduces an additional scalar field
\begin{equation}
\label{eq:pfprofile}
c(x)=1-\exp\left( \frac{-\vert x \vert}{2\ellc}\right)
\end{equation}
with a characteristic length $\ellc$, cf. Figure~\ref{fig:regcrack}, where $c=1$ and $c=0$ resemble intact and fully broken material, respectively. Consequently, a three-dimensional generalization for the corresponding surface energy of the crack~\cite{kuhn_continuum_2010,miehe_thermodynamically_2010}
\begin{equation}
\label{eq:fracener}
\varPsi^\text{c}=\int\limits_{\Omega} 
	\opsi{ }{c}{ }\dv=\int\limits_{\Omega} 
	\frac{\gc}{4\ellc}\left[ \left(1-c\right)^2 +4\ellc^2 \vert\nabla c\vert^2\right]\dv
\end{equation}
can be derived, which is a measure for the energy required to form the corresponding crack surface. The fracture toughness $\gc$ stems from the energetic cracking criterion by Griffith~\cite{griffith_phenomena_1921}. In other words, the crack surface has been regularized~\cite{bourdin_numerical_2000}. In case of a homogeneous fracture toughness, Equation~\eqref{eq:pfprofile} is obtained by rewriting Equation~\eqref{eq:fracener} for the one-dimensional case and subsequent minimization of the functional, subject to the boundary conditions $c(0)=0$ and $c'(x\rightarrow\pm\infty)=0$. For a heterogeneous fracture toughness, phase-field profiles different from Equation~\eqref{eq:pfprofile} are obtained, cf.~\cite{hansen-dorr_phase-field_2020} for more detailed considerations. In Figure~\ref{fig:discif} on the left, a phase-field crack $\gammalc$ is depicted, which possibly interacts with the interface between the two subdomains $\Omega_1$ and $\Omega_2$. Next, the diffuse representation of these subdomains is outlined.
\subsection{Diffuse modeling of elastic heterogeneities}
In the present work, two linear elastic domains $\Omega_i$ with $i=1,2$ are considered. A static order parameter $p\in[0,1]$ describes for every material point whether it belongs to the one $\Omega_1$ or other $\Omega_2$ subdomain, and introduces a smooth transition in the vicinity of the former sharp interface $\gammai$, cf. Figure~\ref{fig:discif}. The transition width is controlled by the interface length scale $\elli$. The order parameter does not change in time and is \textit{a priori} known, e.g. derived from a direct imaging technique. Finally, the motion is described by the displacement $\ve{u}\in\R^3$. The elastic energy density of each domain corresponds to
\begin{equation}
\label{eq:tensSplit1}
\opsi{i}{el}{}= g(c)\,\opsi{i}{el}{+}+\opsi{i}{el}{-}\coma
\end{equation}
where 
\begin{equation}
\label{eq:degradfunc}
g(c)=(1-\eta)\,c^2+\eta
\end{equation}
is a quadratic degradation function to account for previously introduced phase-field cracks. Various other degradation functions are proposed in~\cite{steinke_phase-field_2018,pillai_anisotropic_2020}, including multiple functions for isotropic and anisotropic contributions~\cite{mandal_length_2020} and compared for example in~\cite{kuhn_degradation_2015,strobl_restrictions_2018}. The residual stiffness~$\eta\ll 1$ prevents numerical problems for fully degraded material $c=0$. Furthermore, only the \textit{positive} part $\opsi{i}{el}{+}$ of the strain energy density is degraded to account for the tension compression asymmetry of cracks. Note, that \textit{positive} does not refer to the sign of the strain energy density but to a one-dimensional analogy, where only positive stresses are degraded. A correct and physically meaningful, yet simple and efficient, splitting of the strain energy has been subject of discussion of several publications, cf. for example~\cite{miehe_thermodynamically_2010,amor_regularized_2009,freddi_regularized_2010,strobl_constitutive_2016,steinke_phase-field_2018,bryant_mixed-mode_2018,meng_phase-field_2019,nguyen_implementation_2020}. In case of large deformation kinematics, similar approaches can be found~\cite{borden_isogeometric_2012,hesch_framework_2017,storm_analysis_2019,tang_phase_2019,tarafder_finite_2020}. In this contribution, the so-called tensile split~\cite[Section 3.3]{miehe_thermodynamically_2010} 
\begin{equation}
\label{eq:tensSplit2}
\opsi{i}{el}{\pm} = \frac{\overset{i}{\lambda}}{2}\langle \text{tr}\overset{i}{\eps}\rangle^2_\pm+\overset{i}{\mu}\, \overset{i}{\eps}_\pm:\overset{i}{\eps}_\pm
\end{equation}
with the \textsc{Lamé} constants $\overset{i}{\lambda},\overset{i}{\mu}$ and \textsc{Macaulay} brackets $\langle\bullet\rangle_\pm=\tfrac{1}{2}(\bullet\pm\vert\bullet\vert)$ is adopted. For further information on the strains $\overset{i}{\eps}_\pm$, please refer to the cited section. It is noted, that the diffuse interface model is not restricted to the tensile split, but can be applied to any decomposition of type~\eqref{eq:tensSplit1}. 

The diffuse interface region with $p\in(0,1)$ in the vicinity of the formerly sharp interface $\gammai$ is referred to as $\gammali$. The order parameter $p$, which recovers the values $p=0$ for $\Omega_1$ and $p=1$ for $\Omega_2$ far away from the interface, cf. Figure~\ref{fig:discif}, varies according to
\begin{equation}
\label{eq:orderPar}
p(\ve{x})=\frac{1}{2}\left[\tanh\left(\frac{d(\ve{x})}{\elli}\right)+1\right]\coma
\end{equation}
which corresponds to a three-dimensional generalization of the one-dimensional solution of the \textsc{Modica-Mortola} functional~\cite{modica_esempio_1977}, cf. Figure~\ref{fig:orderparam}. The signed distance $d(\ve{x})$ measures the shortest distance from any point in $\Omega_1\cup\Omega_2$ to the interface mid-surface in the direction of the interface normal $\noi=\nabla p / \vert \nabla p \vert$, cf. Figure~\ref{fig:discif} on the right. In this contribution, the signed distance $d(\ve{x})$, and thus $p(\ve{x})$ are \textit{a priori} known. Within the diffuse region, the elastic energy density is defined as
\begin{equation}
\label{eq:freeenermix}
\psi^\text{el}=(1-p)\,\opsi{1}{el}{}+p\,\opsi{2}{el}{}\coma
\end{equation}
which also holds for the individual phases, because one or the other contribution vanishes for $p=0$ or $p=1$. The approach of interpolating the individual constituents is a widespread approach, cf. for example~\cite{steinbach_multi_2006,schneider_small_2015,schneider_phase-field_2016,mosler_novel_2014,kiefer_numerical_2017}. An alternative representation
\begin{equation}
\psi^\text{el}=g(c)\underbrace{\left[(1-p)\,\overset{1}{\psi}\mathstrut^\text{el}_++p\,\overset{2}{\psi}\mathstrut^\text{el}_+\right]}_{\psi^\text{el}_+}+\underbrace{(1-p)\,\overset{1}{\psi}\mathstrut^\text{el}_-+p\,\overset{2}{\psi}\mathstrut^\text{el}_-}_{\psi^\text{el}_-}
\end{equation}
can be obtained by inserting Equation~\eqref{eq:tensSplit1} into \eqref{eq:freeenermix}. The individual dependencies on $\overset{i}{\eps}$ are omitted above for the sake of readability. Analogously to Equation~\eqref{eq:freeenermix}, an interpolation
\begin{equation}
\label{eq:strainmix}
\eps = (1-p)\,\overset{1}{\eps}+p\,\overset{2}{\eps}
\end{equation}
of the individual strains is assumed in the diffuse region. Together with the strain jump $\jumpeps=\overset{2}{\eps}-\overset{1}{\eps}$, the independent variables for the strain energy density are now changed from $\lbrace\overset{1}{\eps},\overset{2}{\eps} \rbrace$ to $\lbrace \eps,\jumpeps \rbrace$ with the relations
\begin{align}
\label{eq:e1_eeps}
\overset{1}{\eps}&= \eps -p\,\jumpeps\quad\text{and}\\
\label{eq:e2_eeps}
\overset{2}{\eps}&=\eps +(1-p)\,\jumpeps\coma
\end{align}
which is more convenient in view of the presented model. The strain is obtained from the displacement $\ve{u}$ by
\begin{equation}
\eps=\frac{1}{2}\left(\nabla\ve{u}+\left(\nabla\ve{u}\right)\T\right)\coma
\end{equation}
whereas different homogenization assumptions for $\jumpeps$ are discussed in Section~\ref{sec:homass}. The next section combines the energy contributions from the phase-field crack and the linear elastic deformation, and derives the coupled field problem. 
\subsection{Global incremental potential and governing differential equations}
\label{sec:diffeqcdu}
Starting from the findings and definitions of the previous section, the internal energy density 
\begin{equation}
\label{eq:intendens}
\psi\left(\eps,\jumpeps,c\right)=\psi^\text{el} + \psi^\text{c} = g(c)\,\opsi{}{el}{+}\left(\eps,\jumpeps\right) +\opsi{}{el}{-}\left(\eps,\jumpeps\right)+\frac{\mathcal{\gc}}{4\ellc}\left[\left(1-c\right)^2+4\ellc^2\vert\nabla c\vert^2\right]
\end{equation}
is presented, which is used in the following derivations of the governing differential equations. It contains bulk material and crack phase-field contributions, $\psi^\text{el}$ and $\psi^\text{c}$ respectively. The global incremental potential 
\begin{equation}
\label{eq:incrempot}
\Pi^\tau = \int_{\Omega}\psi\!\left(\eps,\jumpeps^\star,c\right)-\psi\!\left(\eps_n,\jumpeps^\star_n,c_n\right)+\tau\,\Phi\!\left(\frac{c-c_n}{\tau}\right)\dv-\int_{\partial\Omega_{\ve{t}}} \bar{\ve{t}}\cdot (\ve{u}-\ve{u}_n)\da
\end{equation}
is defined according to~\cite[Equation (90)]{miehe_multi-field_2011}. The index $n$ refers to the previous, converged time increment, whereas quantities without the index are related to the current $(n+1)$th increment. The time increment is denoted by $\tau=t_{n+1}-t_n$. It is noted, that a first order \textsc{Euler} backward time integration scheme is employed for the rate $\dot{c}$ of the crack phase-field, which enters the dissipation potential $\Phi(\dot{c})=\frac{\eta_\text{f}}{2}\dot{c}^2$. Here, $\etaf\geq 0$ is a kinetic parameter, often referred to as viscosity, which is of purely numerical nature in this contribution. Its choice is discussed in Section~\ref{sec:numex}. The incorporation of $\dot{c}$ can be understood as a viscous regularization, which makes it possible to use a monolithic solution approach. Various other approaches and their implications can be found in literature, see e.g.~\cite{miehe_phase_2010,gerasimov_line_2016}. The vector $\bar{\ve{t}}$ denotes given external tractions on \textsc{Neumann} boundaries $\partial\Omega_{\ve{t}}$. The strain jump, which enters $\psi$, is determined within a local minimization procedure, cf. Section~\ref{sec:homass}, and is therefore known which is indicated by the star $\jumpeps^\star$. Hence, $\ve{u}$ and $c$ are obtained by the global minimization principle 
\begin{equation}
\lbrace \ve{u} ,c\rbrace = \arg\inf_{\ve{u}\in\mathcal{U}}\inf_{c\in\mathcal{C}}\Pi^\tau
\end{equation}
with admissible sets
\begin{align}
\mathcal{U}(\ve{u}) &= \lbrace \ve{u}\in \mathbb{R}^3 \;\vert\; \ve{u} = \bar{\ve{u}} \enspace \text{on} \enspace \partial\Omega_{\ve{u}}\rbrace\quad\text{and} \\
\mathcal{C}(c) &= \lbrace c \in \left[0,1\right] \subset \mathbb{R} \;\vert\; c=0 \enspace\text{on}\enspace\gammac_0 \enspace \text{and}\enspace c({\ve{x}})=0\enspace \text{for}\enspace c_n({\ve{x}})<c_\text{th} \rbrace \point
\end{align}
The first constraint in $\mathcal{C}$ is only relevant for the initial crack along $\gammac_0$. Afterwards, the second condition, also known as \textit{fracture-like} irreversibility constraint~\cite{kuhn_continuum_2010,linse_convergence_2017}, suffices, where $c_\text{th}\ll 1$ is a small, positive threshold value. Previous investigations~\cite{hansen-dorr_phase-field_2020} confirmed, that setting $c_\text{th}=0.03$ does not impact the crack path, while ensuring irreversibility. A discussion on different irreversibility constraints can be found in~\cite{linse_convergence_2017}. Subsequently, the \textsc{Euler-Lagrange} equations 
\begin{align}
\label{eq:linmom}
\nabla\cdot\sig &=\bold{0}\quad\text{and}\\ \label{eq:pfevo}
\eta_\text{f}\,\frac{c-c_n}{\tau} &=\frac{\gc}{2\ellc}(1-c)-2(1-\eta)\,c\,\psi^\text{el}_++2\ellc \nabla\cdot(\gc \nabla c)
\end{align}
can be derived, subject to \textsc{Neumann} boundary conditions
\begin{align}
\sig\T\cdot\nob&=\bar{\ve{t}}\quad \text{on}\quad \partial\Omega_{\ve{t}}\quad\text{and}\\
\nabla c\cdot\nob&=0\quad \text{on}\quad \partial\Omega\coma
\end{align} 
where $\nob$ denotes the outward normal vector along the corresponding boundary. Equation~\eqref{eq:degradfunc} has been used in Equation~\eqref{eq:pfevo}. The thermodynamically consistent relation for the \textsc{Cauchy} stress tensor is defined by
\begin{equation}
\label{eq:cauchystress}
\osig{ }{ }=\frac{\partial \opsi{ }{el}{ }}{\partial\eps} = g(c)\left((1-p)\,\osig{1}{+}+p\,\osig{2}{+}\right)+(1-p)\,\osig{1}{-}+p\,\osig{2}{-}\quad\text{with}\quad\osig{i}{\pm} = \partial \opsi{i}{el}{\pm}\big/\partial\overset{i}{\eps}\point
\end{equation}
It is noted, that the chain rule has been used together with Equations~\eqref{eq:e1_eeps} and~\eqref{eq:e2_eeps}. The next section is dedicated to the determination of the strain jump $\jumpeps^\star$ which enters the incremental potential~\eqref{eq:incrempot}. The validity of Equation~\eqref{eq:cauchystress} is demonstrated in Appendix~\ref{app:stress}.

\subsection{Homogenization assumptions within the diffuse interface region}
\label{sec:homass}
In this section, two different approaches from the homogenization theory are examined to determine the unknown strain jump $\jumpeps$, which enters the incremental potential~\eqref{eq:incrempot} as $\jumpeps^\star$. In particular, the Voigt-Taylor approach and a partial rank-I relaxation are introduced. In contrast to existing works~\cite{schneider_phase-field_2016}, the combination of an additive strain energy decomposition and the above approaches is presented. This work incorporates the tensile split by Miehe~\cite{miehe_thermodynamically_2010} but is not limited to it. 
\paragraph{Voigt-Taylor approach:}
In classical homogenization theory, the Voigt-Taylor approach assumes equal strains in every constituent, i.e.
\begin{equation}
\overset{1}{\eps}=\overset{2}{\eps}=\eps\point
\end{equation}
From that, the strain jump $\jumpeps^\star=\bold{0}$ follows, which is inserted in the incremental potential above. In other words, the Voigt-Taylor approach is automatically implemented in any diffuse modeling framework similar to this one, if the strain jump is not treated by some means or other. 

Different from the assumption of equal strains in the constituents, the Reuss-Sachs approach assumes equal stresses. It is, however, not investigated in this contribution, because there are various other publications doing so~\cite{mosler_novel_2014,schneider_phase-field_2015,kiefer_numerical_2017}. The two approaches yield upper and lower energetic barriers for the energy, respectively, cf.~\cite[Equation~(55)]{mosler_novel_2014} or \cite[Equation~(62)]{kiefer_numerical_2017}.

\paragraph{Partial rank-I relaxation:} The previous approaches do not include any information about the interface but only interpolate between the bulk energies, cf. Equation~\eqref{eq:freeenermix}. In contrast, the partial rank-I relaxation correctly incorporates all mechanical jump conditions by means of the strain jump, which can be expressed by the \textsc{Hadamard} condition~\cite[Equation~(2.2.9)]{silhavy_mechanics_1997}
\begin{equation}
\label{eq:strjumppr1}
\jumpeps=\frac{1}{2}\left(\ve{a}\otimes\noi +\noi\otimes\ve{a}  \right)
\end{equation}
without loss of generality\footnote{The \textsc{Hadamard} condition holds for a continuous deformation field and is also referred to as kinematical compatibility condition. A proof is provided in~\cite[Sec. 2.1.6]{silhavy_mechanics_1997}. Because of the regularized description of the crack, the displacement field is always continuous. Thus, the condition also holds when a crack meets the interface.}, and the stress jump $\jump{\sig}\T\cdot\noi=\bold{0}$. The former is stated here for a symmetric tensor $\eps$. The vector $\ve{a}$ is the strain jump amplitude and $\noi = \nabla p /\vert\nabla p  \vert$ the interface normal. The latter is fulfilled as follows: In order to calculate $\ve{a}$, a pointwise minimization procedure~\cite[Equation~(50)]{mosler_novel_2014}
\begin{equation}
\label{eq:inflocal}
\opsi{}{el}{}\left(\eps,\jumpeps^\star,c\right)=\inf_{\ve{a}\in\mathbb{R}^3}\opsi{}{el}{}\left(\eps,\jumpeps,c\right)
\end{equation}
with the associated necessary condition
\begin{equation}
\label{eq:neccondpr1}
0\overset{!}{=}p(1-p)\left\lbrace \left[g(c)\left(\osig{2}{+}-\osig{1}{+}\right)+\osig{2}{-}-\osig{1}{-}\right]\T\cdot\noi\right\rbrace
\end{equation}
is pursued, which is a relaxation of $\opsi{ }{el}{ }$ with respect to $\ve{a}$~\cite{kiefer_numerical_2017}. The stresses $\osig{i}{\pm}$ are calculated according to Equation~\eqref{eq:cauchystress}. For the cases $p=0$ or $p=1$, which indicate bulk material away from the diffuse interface region, the condition is fulfilled and the jump $\jumpeps=\bold{0}$ either way. For $p\in(0,1)$, the expression in the curly braces has to be zero, which is similar to the aforementioned stress jump condition at sharp interfaces. Here, a pointwise enforcement throughout the whole diffuse interface region is required. In absence of a crack, $g(1)=1$, the analytical solution for linear elastic material according to \cite[Equation (42)]{kiefer_numerical_2017} is recovered because the additive strain energy decomposition is unnecessary. The same is valid for $g(c)<1$ if there is no strain energy decomposition at all, cf. \cite{schneider_phase-field_2016}, because $g(c)$ does not exhibit any jump across the interface. In this contribution, the key novelty lies in the incorporation of any (variational) additive decomposition of the strain energy into a degraded and persistent part. The nonlinearity of the spectral decomposition for the tensile split requires a local \textsc{Newton-Raphson} scheme to solve for $\ve{a}$ as soon as a crack emerges within the diffuse interface region. In the context of an elasto-plastic multiphase-field model, Schneider et al.~\cite{schneider_small_2015} had to solve a very similar equation to \eqref{eq:neccondpr1} because of the nonlinearity due to plasticity. The local \textsc{Newton-Raphson} scheme is presented in Appendix~\ref{app:localnr}.

\subsection{Numerical implementation}
\label{sec:numimpl}
\begin{figure*}[t]
	\centering
	\includegraphics{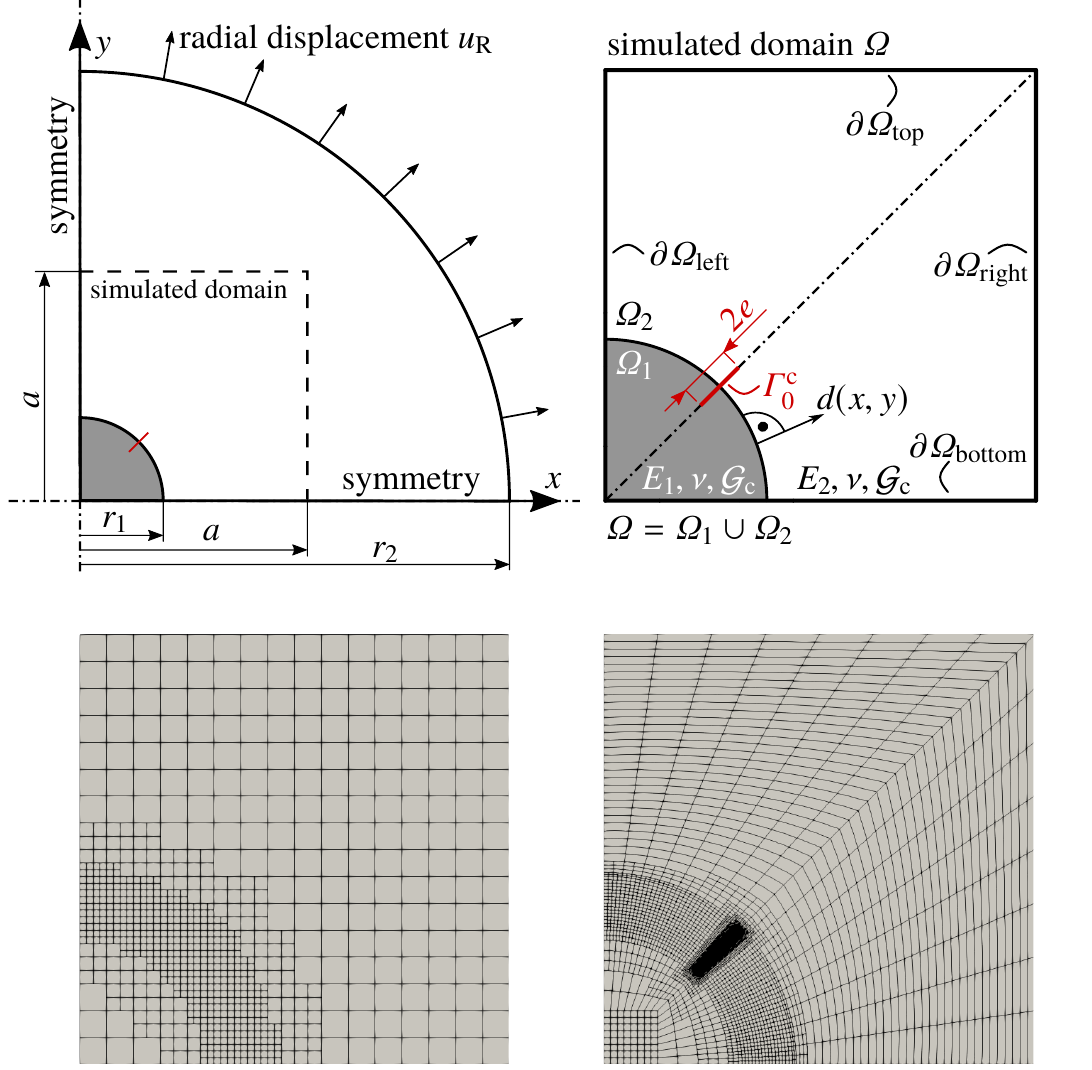}
	\caption{Setup for two-dimensional convergence study: \textbf{Top left}, the circular domain with a circular inclusion is depicted. A radial displacement is applied to the outer boundary. For the simulation, only the dashed square region is considered and depicted \textbf{top right}. Two studies, without and with the crack of length $2e$, are conducted. The associated meshes are presented in the \textbf{bottom} row. Only the upmost refinement level of the NURBS meshes is shown. 
	}
	\label{fig:circInclGeom}
\end{figure*}

The weak form including consistent linearization for the global \textsc{Newton-Raphson} scheme can be found in Appendix~\ref{app:weaklinear}. The spatial discretization is carried out using locally refined Truncated Hierarchical non-uniform rational B-splines~\cite{hennig_bezier_2016} (NURBS) with quadratic shape functions in Section~\ref{sec:convStudy} and linear shape functions in Section~\ref{sec:crackProp}. An adaptive refinement strategy is pursued, which allows for efficient computations with a high resolution of the steep gradient in regions where the crack develops and propagates. Additionally, the mesh is prerefined along the interface $\gammali$ to resolve the diffuse transition for $p$ sufficiently. Spatial convergence with respect to the crack length scale $\ellc$ is ensured by choosing the finest element level in a way, that the characteristic element size is at least three times smaller than $\ellc$, cf. Figure~\ref{fig:circInclGeom}. All simulations are carried out using a Matlab-based in-house finite element code. 

\section{Numerical examples}
\label{sec:numex}

\subsection{Convergence study}
\label{sec:convStudy}
For verification of the implementation, the convergence of the presented modeling approach with respect to the sharp interface solution is tested. For this purpose, a quarter of a circular bi-material disc is subjected to a radial displacement $u_\text{R}=r_2/1000$, cf. Figure~\ref{fig:circInclGeom}. In the study, the simulated domain is reduced to a square. The geometry parameters are given as $r_1=\SI{3}{\milli\meter}$,  $r_2=\SI{15}{\milli\meter}$ and  $a=\SI{8}{\milli\meter}$. Plane strain is assumed. The signed distance for Equation~\eqref{eq:orderPar} reads
\begin{equation}
d(\ve{x})=d(x,y)=\sqrt{x^2+y^2}-r_1\point
\end{equation} 
Symmetry boundary conditions are applied along $\partial\Omega_\text{left}$ and $\partial\Omega_\text{bottom}$. Along $\partial\Omega_\text{top}$ and $\partial\Omega_\text{right}$, a traction $\bar{\ve{t}}$ is applied, cf. Equation~\eqref{appeq:traction} in Appendix~\ref{app:traction}. Two error norms, the well known energy norm
\begin{equation}
e_\text{loc} = \frac{\int \vert(\sig-\sig_\text{ex}):(\eps-\eps_\text{ex})\vert\,\text{d}\Omega}{\int \sig_\text{ex}:\eps_\text{ex}\,\text{d}\Omega}
\end{equation}
and total error in energy
\begin{equation}
e_\text{tot} = \left\vert\frac{\tfrac{1}{2}\int \sig:\eps\,\text{d}\Omega}{\tfrac{1}{2}\int \sig_\text{ex}:\eps_\text{ex}\,\text{d}\Omega}-1\right\vert
\end{equation}
are defined, where the index $(\bullet)_\text{ex}$ stands for the exact solution and $\vert\bullet\vert$ stands for the absolute value. The subscript `loc' for the energy norm $e_\text{loc}$ refers to the fact, that the error is calculated locally before integration. For the case without a crack, the analytical solution, cf. Appendix~\ref{app:traction}, is used for $(\bullet)_\text{ex}$, and the denominator in $e_\text{tot}$ is replaced by the analytical expression~\eqref{appeq:energ}. For the cracked case, an overkill solution with a sharp interface serves as `exact' solution. The overkill solution is calculated on a much finer mesh. The evaluation of the energy norm $e_\text{loc}$ is not possible because of differing integration point locations. 

\paragraph{Investigation without a crack}
At first, the model is tested for the linear-elastic case in the absence of a crack. The material parameters are set to $\nu_1=\nu_2=\nu=0.3$, $E_2=\SI{100}{\giga\pascal}$, $\ellc=\SI{50}{\micro\meter}$ and a very high value $\gc=\SI{e5}{\kilo\newton\per\milli\meter}$ to prevent any crack formation or propagation. The viscosity $\etaf$ is set to zero. Two cases $E_1/E_2=\lbrace2,20\rbrace$ allow to study the influence of the intensity of the elastic dissimilarity. The coarsest element level of the mesh is depicted in Figure~\ref{fig:circInclGeom} on the bottom left. An $h\elli$-refinement study is undertaken, where the interface length scale and the mesh size are reduced simultaneously, i.e. more refinement levels are added along the arc with radius $r_1$. The ratio $\elli/h\approx2.7$ is kept constant.

\begin{figure}[t]
	
	\centering
	\subfloat[Energy norm $e_\text{loc}$ for $E_1/E_2=\lbrace 2,20\rbrace$]{
		\label{fig:noCrack2D_ld_en}
		\includegraphics[trim={0.0cm 0cm 0.2cm 0cm},clip]{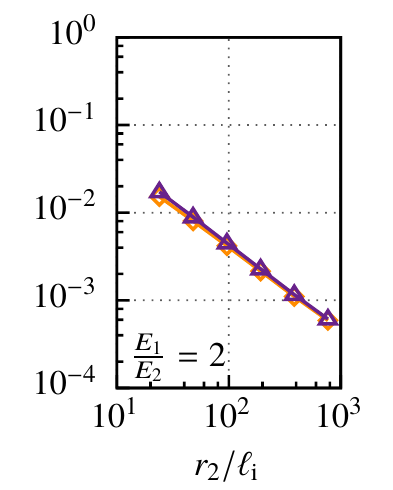}
		\includegraphics[trim={0.9cm 0cm 0cm 0cm},clip]{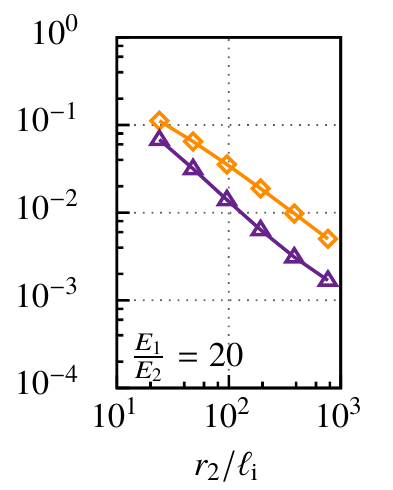}
	}
	\subfloat[Total error in energy $e_\text{tot}$ for $E_1/E_2=\lbrace 2,20\rbrace$]{
		\label{fig:noCrack2D_ten}
		\includegraphics[trim={0.0cm 0cm 0.2cm 0cm},clip]{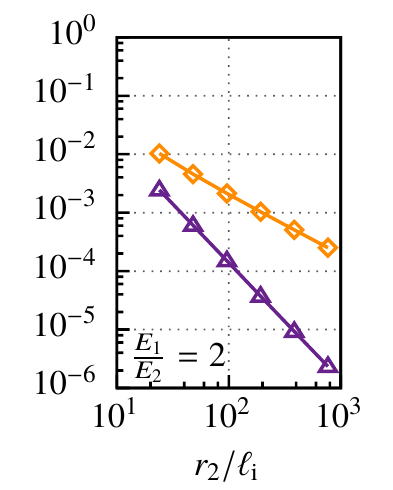}
		\includegraphics[trim={0.9cm 0cm -0.0cm 0cm},clip]{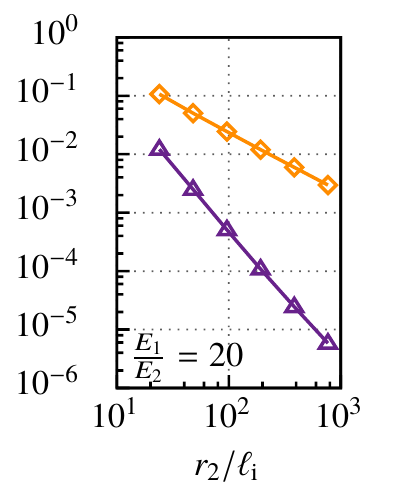}
	}
	\\
	\vspace{0.3cm}
	\footnotesize 
	\begin{tabular}{l|l}
		\legline{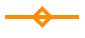} Voigt-Taylor & \legline{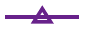} partial rank-I \\   
	\end{tabular}
	\caption{The local (a) and total (b) error in energy for the convergence investigation in the absence of a crack are presented. First, the partial rank-I scheme exhibits the same or better convergence rates and error levels. Second, the Voigt-Taylor approach reacts more sensitive to a change in the elastic dissimilarity, which can be seen from a comparison between the left and the right plot in (a) and~(b). 
	}
	\label{fig:noCrack2D}
\end{figure}

The results of the convergence study are depicted in Figure~\ref{fig:noCrack2D}. In general, it can be seen, that the partial rank-I scheme exhibits the same or better convergence rates and error levels than the Voigt-Taylor approach with respect to the analytical sharp interface solution. Furthermore, the obtained results are consistent with literature, compare Figure~\ref{fig:noCrack2D_ld_en} to~\cite[Figure 8b]{hennig_diffuse_2019}, or Figure~\ref{fig:noCrack2D_ten} to~\cite[Figure 5b]{kiefer_numerical_2017}. Additionally, the Voigt-Taylor approach reacts more sensitive to a more intense elastic dissimilarity which can be seen when comparing the left and right plots in Figures~\ref{fig:noCrack2D_ld_en} and \ref{fig:noCrack2D_ten}. This is also reported in~\cite{steinbach_multi_2006}.

\paragraph{Investigation with a crack across the interface}
Next, a phase-field crack along $\gammac_0$ with $e=\SI{0.5}{\milli\meter}$ is introduced as shown in Figure~\ref{fig:circInclGeom}, with the corresponding finite element mesh on the bottom right. The mesh is prerefined to the finest level along the crack to rule out any inaccuracies due to approximation errors of the phase-field. It serves as coarsest refinement level and is gradually refined along the interface for the $h\elli$-convergence study. All parameters and the boundary conditions are chosen similar to the previous investigation, except that only results for $E_1/E_2=2$ are considered for the sake of brevity. It is noted, that the traction $\bar{\ve{t}}$ does not reproduce the analytical solution as before, nor does it lead to a circumferentially constant radial displacement $u_\text{R}$, which is why an overkill solution is used as reference. For the overkill solution, the NURBS mesh has to account for the physical $C^0$-continuity of the displacement along the arc $r_1$. Additionally, the crack boundary condition along $\gammac_0$ has to be set. This is why the finite element mesh is different from the mesh of the previous convergence study. Convergence of the overkill solution has been verified. 

\begin{figure}[t]
	\centering
	\subfloat[$\eta = 10^{-2}$]{
		\label{fig:crack2D_0.01}
		\includegraphics[trim={0.0cm 0cm 0.2cm 0cm},clip]{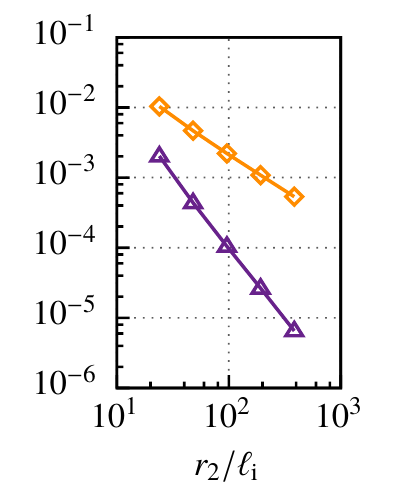}
	}
	\subfloat[$\eta = 10^{-5}$]{
		\label{fig:crack2D_0.00001}
		\includegraphics[trim={0.9cm 0cm 0.2cm 0cm},clip]{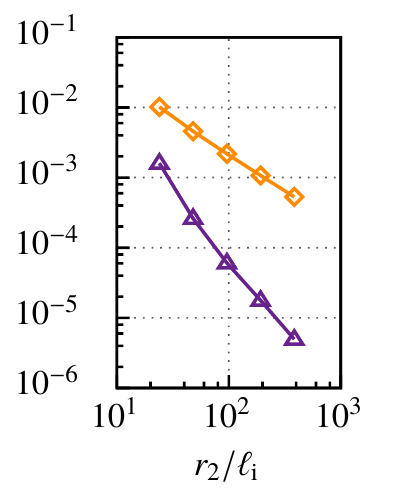}
	}
	\subfloat[$\eta = 0$]{
	\label{fig:crack2D_0.0}
	\includegraphics[trim={0.9cm 0cm 0.2cm 0cm},clip]{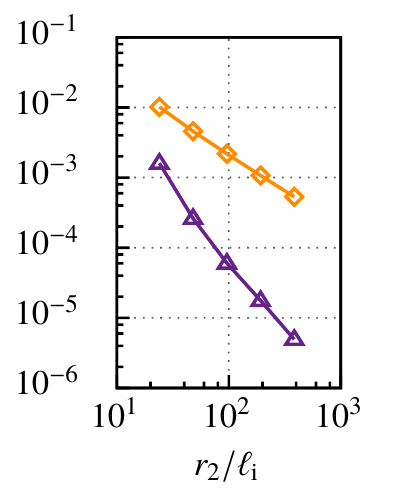}
	}
	\\
	\vspace{0.3cm}
	\footnotesize 
	\begin{tabular}{l|l}
		\legline{22} Voigt-Taylor & \legline{23} partial rank-I \\   
	\end{tabular}
	\caption{The total error in energy $e_\text{tot}$ for the convergence investigation with a crack across the interface is presented for various residual stiffnesses (a) to (c). The residual stiffness does not have an influence on the convergence behavior. Furthermore, the convergence rates and error levels are similar to the case without a crack. 
	}
	\label{fig:diffif_crack}
\end{figure}

The results for three different values of the residual stiffness $\eta$, cf. Equation~\eqref{eq:degradfunc}, are depicted in Figure~\ref{fig:diffif_crack}. All plots are identical, i.e. the residual stiffness does not influence the convergence, which is expected. However, for crack propagation simulations, a zero value most probably leads to numerical problems. Furthermore, the obtained results are more or less identical to the intact case, cf. Figure~\ref{fig:diffif_crack} to the left plot in Figure~\ref{fig:noCrack2D_ten}: Convergence rate and error level are matching. 

The two convergence studies suggest the successful verification of the implemented method and validation by means of literature results. The major question whether any impacts on propagating cracks can be observed is answered in the next section.

\subsection{Crack branching at interfaces}
\label{sec:crackProp}
\begin{figure}[t]
	\centering
	\raisebox{-0.9cm}{\includegraphics{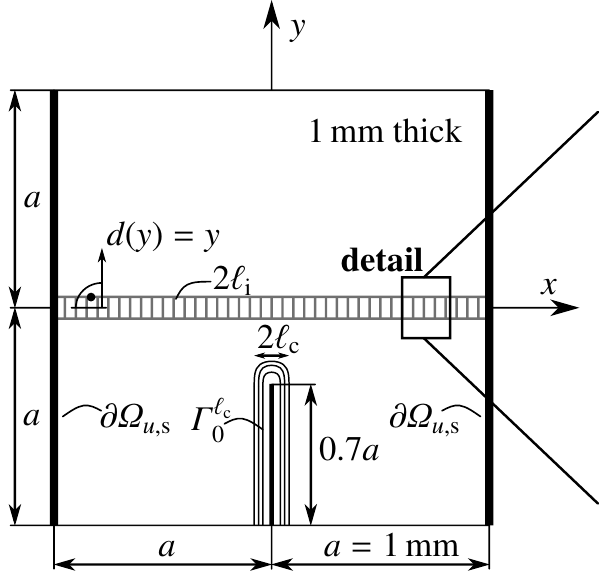}}\hspace{-0.4cm}
	\includegraphics[angle=90,origin=c]{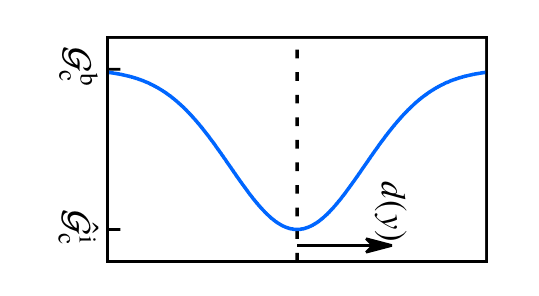}\hspace{-0.4cm}
	\includegraphics[angle=90,origin=c]{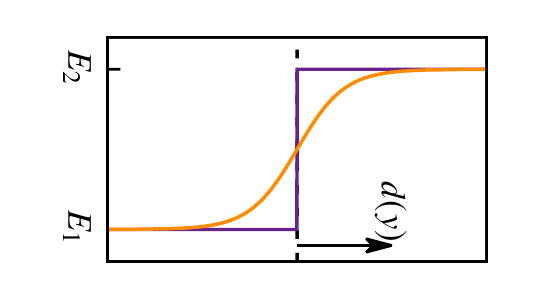}
	\caption{The general setup for the crack branching study is sketched on the left. The diffuse interface is schematically depicted by the grey hatched region.	The material parameters are assigned according to the functions depicted on the right, using values of $E_1$, $E_2$, $\hat{\mathcal{G}}_\text{c}^\text{i}$ and $\gcb$. 
		The initial phase-field crack is depicted by $\gammalc_0$. \textsc{Dirichlet} boundary conditions are applied for the displacement along the bold marked side edges $\partial\Omega_{u,\text{s}}$ whereas homogeneous \textsc{Neumann} boundary conditions are applied on $\partial\varOmega\setminus \partial\varOmega_{u,\text{s}}$. 
	}
	\label{fig:simsetup}
\end{figure}

\begin{figure}[t]
	\centering
	\subfloat[Prerefined mesh]{
		\label{fig:meshBranch}
		\includegraphics[trim={0cm 5cm 5cm 0cm},clip]{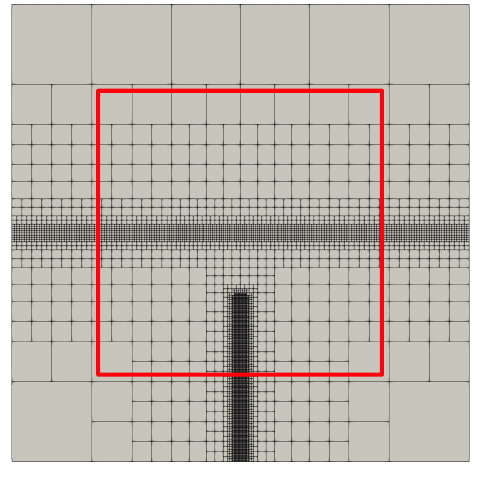}
	}
	\subfloat[Sharp elastic discontinuity]{
		\label{fig:resSharp}
		\includegraphics[trim={5cm 5cm 0cm 0cm},clip]{fig7abcd.pdf}
	}\hspace*{1.2cm}~\\
	\subfloat[Diffuse Voigt-Taylor]{
		\label{fig:resVT}
		\includegraphics[trim={0cm 0cm 5cm 5cm},clip]{fig7abcd.pdf}
	}
	\subfloat[Diffuse partial rank-I]{
		\label{fig:resPR1}
		\includegraphics[trim={5cm 0cm 0cm 5cm},clip]{fig7abcd.pdf}
	}\hspace{0.3cm}
\raisebox{0.73cm}{\includegraphics{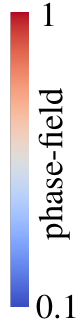}}
	\caption{In (a), the NURBS mesh is depicted. The red square marks the location of the contours in (b) to (d). In total, seven hierarchical levels are used. The simulation results for $\alpha=-0.5$ and $\gcb/\gcih=23$ are depicted in (b) to (d). For these three plots, the two black lines mark the interface midline with $d\equiv0$. Shortly after meeting the interface, branching is predicted by the sharp reference solution. In case of a diffuse interface, the partial rank-I relaxation predicts the same phenomenon, while the Voigt-Taylor approach fails. The visibility of elements with $c<0.1$ is turned of which is why the contour legend stops at 0.1. }
	\label{fig:meshContour}
\end{figure}
Besides the improved convergence of the presented model with respect to the sharp interface solution it is of great interest whether any impacts on crack propagation at diffuse interfaces are recognized. Hansen-Dörr et al.~\cite{hansen-dorr_phase-field_2020} thoroughly investigated many different setups, where a crack approaches a possibly inclined interface and, depending on the ratio of the elastic moduli on both sides of the interface and the ratio of the bulk material and interface fracture toughnesses, decides to branch or to go straight into the adjacent bulk material. A comparison to analytical results from He and Hutchinson~\cite{he_crack_1989} and to a sharp interface reference revealed inaccuracies of the Voigt-Taylor approach, which was used therein. This section demonstrates, that the presented model extension to a partial rank-I relaxation yields the same critical deflection ratios as the sharp interface reference, while the Voigt-Taylor gives different results.

The setup for this study, which is similar to~\cite{hansen-dorr_phase-field_2020}, is depicted in Figure~\ref{fig:simsetup}. The corresponding NURBS mesh is shown in Figure~\ref{fig:meshBranch}. A square domain with an initial phase-field crack $\gammalc_0$ is subjected to a horizontal displacement load. For this purpose, the concept of the so-called \textit{surfing boundary condition} \cite{hossain_effective_2014} is exploited, which proved to be very suitable for quantification purposes~\cite{hossain_effective_2014,kuhn_discussion_2016}.
The key idea of this approach is to introduce a point with the time dependent position~$\bar{\ve{x}}$, which is moving along the $y$-axis
\begin{equation}
\bar{\ve{x}}
= \begin{bmatrix}  \bar{x} \\ \bar{y}  \end{bmatrix}=
\begin{bmatrix}  0 \\ v \cdot t + \bar{y}_0  \end{bmatrix}
\point
\end{equation}
A displacement of hyperbolic tangent-like shape is applied on the side edges $\partial \varOmega_{u, \text{s}}$,
\begin{equation}
\bar{\ve{u}} =
\frac{u_\nv{ref}}{2} \, \left( 1- \tanh \left[ \frac{y-\bar{y}}{d}\right] \right) \, 
\begin{bmatrix}  \sign ( x) \\ 0  \end{bmatrix}
\coma
\label{eq:sbc-tanh}
\end{equation}
with respect to the moving point, assuming $u_\text{ref} = \SI{7.5}{\micro\meter}$, $d=\SI{0.5}{\milli\meter}$, $v=\SI{0.3}{\milli\meter\per\second}$ and $\bar{y}_0=\SI{-4}{\milli\meter}$. Similar parameters proved to be suitable in previous studies~\cite{hansen-dorr_phase-field_2020}. Homogeneous \textsc{Neumann} boundary conditions are considered along the top and bottom edges. 
A horizontal interface, implied by the grey hatched bar in Figure~\ref{fig:simsetup}, divides the domain in two halves. The fracture toughness of the interface is different from the bulk material. This is accounted for by locally reducing the fracture toughness according to a \textsc{Gaussian}-like distribution
\begin{equation}
\label{eq:rgauss}
\rgauss(d(y),\gcb,\gcih)=\gcb-\left( \gcb-\gcih \right) \,\exp\left[-\left(\frac{d(y)}{2\elli}\right)^2\right]\coma
\end{equation}
cf. Figure~\ref{fig:simsetup} blue plot in the middle. Far away from the interface, the bulk material fracture toughness $\gcb=\SI{2.7}{\newton\per\milli\meter}$ is recovered. Along the interface, a potential crack is supposed to propagate for a critical energy release rate equal to $\gci$. However, the crack experiences a bulk material influence due to the interaction of both length scales $\elli$ and $\ellc$~\cite{hansen-dorr_phase-field_2020}, and tends to propagate at critical energy release rates, which are higher compared to the minimum of the function in Equation~\eqref{eq:rgauss}. Hence, a compensated interface fracture toughness $\gcih$ is specified in a way, that a crack propagates along the interface midline at the true interface fracture toughness~$\gci$, i.e. $\gcih<\gci$. For more details on the compensation procedure, the reader is referred to~\cite{hansen-dorr_phase-field_2020,hansen-dorr_phase-field_2019,hansen-dorr_numerical_2017}. The ratio of the elastic moduli with $E_1=\SI{210}{\giga\pascal}$ varies, while the \textsc{Poisson} ratio $\nu=0.3$ is kept constant, i.e. the interpolation in Equation~\eqref{eq:freeenermix} directly translates into a hyperbolic tangent function for $E_1$ and $E_2$, cf. orange line in Figure~\ref{fig:simsetup} on the right. For the sharp interface reference solution, the elastic moduli vary according to the purple step function. The elastic dissimilarity is described with the first \textsc{Dundurs}' parameter~\cite{dundurs_discussion:_1969}
\begin{equation}
\alpha = \frac{E_2-E_1}{E_1+E_2}\point
\end{equation}
Following the study of Hansen-Dörr et al.~\cite{hansen-dorr_phase-field_2020}, the length scale parameters are set to $\elli=\SI{18.75}{\micro\meter}$ and $\ellc=\SI{15}{\micro\meter}$, while the viscosity and the residual stiffness take the values $\etaf=\SI{e-5}{\kilo\newton\second\per\square\milli\meter}$ and $\eta=\SI{e-5}{}$, respectively.

A parameter study with $\alpha=\lbrace -0.5,-0.25,0.25,0.5 \rbrace$ and different ratios $\gcb/\gci$ is conducted, and the results for the sharp interface and the diffuse interface description using the Voigt-Taylor approach or partial rank-I relaxation are compared with respect to the crack behavior at the interface: Fixing $\alpha$, the tendency of the crack to branch is higher for larger ratios $\gcb/\gci$. The NURBS mesh is adaptively refined according to the crack path. All elements, where the phase-field fell below a value of $c=0.5$, are marked for refinement~\cite{hennig_bezier_2016}. For the sake of brevity, only the results where the two homogenization approaches yielded different crack phenomena at the interface are presented.

Figures~\ref{fig:resSharp}--\ref{fig:resPR1} present the crack patterns shortly after the crack met the interface, with $\alpha=-0.5$ and $\gcb/\gci=8$, i.e. $\gcb/\gcih=23$. For better interpretation, the elements with $c<0.1$ are made invisible. It can be seen, that the sharp interface solution and the diffuse interface approach with a partial rank-I relaxation predict a branching of the crack. On the contrary, the Voigt-Taylor approach yields a straight crack with no such tendency. Even setting $\gcb/\gci=10$, i.e. $\gcb/\gcih=30.3$, did not change this fact. This is in line with the presumption of Hansen-Dörr et al.~\cite{hansen-dorr_phase-field_2020}, who also observed such deviations. For further loading, all cracks started to propagate further in $y$-direction, while the two crack branches for the sharp interface and the diffuse partial \mbox{rank-I} relaxation arrested. For $\alpha=-0.25$ and $\gcb/\gci=5.5$, i.e. $\gcb/\gcih=12$, a similar behavior is observed. For $\alpha=\lbrace 0.25,0.5 \rbrace$, the three approaches always yield the same outcome. On the one hand this could imply an increased sensitivity of the crack behavior towards the homogenization scheme for negative $\alpha$ values. On the other hand, only discrete values of $\gcb/\gci$ are investigated and it is possible, that only those were picked from the infinite amount of possible ratios, where the results are identical. This is, why future model investigations will include a deeper and also three-dimensional analysis of the phenomena reported above. To conclude with, it was shown, that the present model extension meaningfully influences the crack behavior and yields closer results to the sharp interface solution. 

\section{Conclusions}
\label{sec:conclusion}
In this contribution, a variational diffuse modeling approach was presented. Heterogeneities were described by an order parameter which is obtained from the signed distance function and a hyperbolic tangent, and smoothly bridges material discontinuities. The mechanical boundary value problem is monolithically coupled to a phase-field model for cracks. 

Due to the diffuse transition between two materials, the mechanical jump conditions are not necessarily fulfilled for a general case. Instead, without taking any measures, the Voigt-Taylor homogenization approach is implemented within the diffuse region, which states equal strains in both phases. As shown above, this fails to reproduce several crack patterns compared to the sharp interface. A remedy to this issue is the partial rank-I relaxation, which was extended to account for a strain energy density decomposition. The decomposition is in general necessary for physically meaningful crack behavior like closure. The partial rank-I relaxation yields a point-wise fulfillment of the mechanical jump conditions and was successfully validated by means of convergence investigations. The presented scheme includes a local \textsc{Newton-Raphson} loop, which is the main drawback of the approach: Although, the local algorithm always converged without any problems, it is more time consuming than avoiding it by employing the Voigt-Taylor approach. A final study of a crack propagating towards an interface demonstrated, that the partial rank-I relaxation does not only improve the convergence of the global energy, but also influences the crack growth locally towards the sharp interface solution since it alters the crack driving force directly. The Voigt-Taylor approach did not achieve this for the given length scale of the interface, which is in line with observations from literature, and underlines the necessity to fulfill the mechanical jump conditions. 

\begin{acknowledgements}
The authors gratefully acknowledge support by the Deutsche Forschungsgemeinschaft in the Priority Program 1748 “Reliable simulation techniques in solid mechanics. Development of non-standard discretization methods, mechanical and mathematical analysis” under the project KA3309/3-2. The computations were performed on a HPC-Cluster at the Center for Information Services and High Performance Computing (ZIH) at TU Dresden. The authors thank the ZIH for allocations of computational time.
\end{acknowledgements}
\begin{small}
\textbf{Conflict of Interest}~~The authors declare that they have no conflict of interest.
\end{small}

\appendix
\section*{Appendix}
\section{-- Stress-strain relationship}
\label{app:stress}
In this section, the validity of Equation~\eqref{eq:cauchystress} is demonstrated. For better readability of the tensor products, the index notation is used where appropriate. Differentiation of Equation~\eqref{eq:freeenermix} with respect to the strain yields
\begin{equation}
\label{appeq:psieleps}
\frac{\partial\opsi{ }{el}{ }}{\partial \varepsilon_{lm}} = (1-p) \frac{\partial\opsi{1}{el}{ }}{\partial \overset{1}{\varepsilon}_{ij}} \frac{\partial\overset{1}{\varepsilon}_{ij}}{\partial \varepsilon_{lm}} + p\frac{\partial\opsi{2}{el}{ }}{\partial \overset{2}{\varepsilon}_{ij}} \frac{\partial\overset{2}{\varepsilon}_{ij}}{\partial \varepsilon_{lm}}=(1-p) \, \overset{1}{\sigma}_{ij} \frac{\partial\overset{1}{\varepsilon}_{ij}}{\partial \varepsilon_{lm}} + p\, \overset{2}{\sigma}_{ij} \frac{\partial\overset{2}{\varepsilon}_{ij}}{\partial \varepsilon_{lm}}\coma
\end{equation}
where $\osig{1}{ }$ and $\osig{2}{ }$ correspond to the stresses in the individual phases.
Subsequently, the strains of phase 1 and 2 are differentiated with respect to the interpolated strain using Equations~\eqref{eq:e1_eeps}, \eqref{eq:e2_eeps} and~\eqref{eq:strjumppr1}, and one obtains
\begin{align}
\label{appeq:eps1eps}
\frac{\partial\overset{1}{\varepsilon}_{ij}}{\partial \varepsilon_{lm}}&=\frac{1}{2}\left( \delta_{il}\delta_{jm}+\delta_{im}\delta_{jl}  \right) - p\frac{\partial\jump{\varepsilon_{ij}}}{\partial a_o}\frac{\partial a_o}{\partial \varepsilon_{lm}}\quad\text{and}\\ \label{appeq:eps2eps}
\frac{\partial\overset{2}{\varepsilon}_{ij}}{\partial \varepsilon_{lm}}&=\frac{1}{2}\left( \delta_{il}\delta_{jm}+\delta_{im}\delta_{jl}  \right) + (1-p)\frac{\partial\jump{\varepsilon_{ij}}}{\partial a_o}\frac{\partial a_o}{\partial \varepsilon_{lm}}\coma
\end{align}
where $\delta_{ij}$ denotes the \textsc{Kronecker} symbol. Inserting Equations~\eqref{appeq:eps1eps} and \eqref{appeq:eps2eps} into Equation~\eqref{appeq:psieleps} yields
\begin{equation}
\label{appeq:stressstrain}
\frac{\partial\opsi{ }{el}{ }}{\partial \varepsilon_{lm}} = (1-p) \,\overset{1}{\sigma}_{lm} + p\,\overset{2}{\sigma}_{lm}
+ p\,(1-p)\left[\overset{2}{\sigma}_{ij}-\overset{1}{\sigma}_{ij}   \right]\frac{\partial\jump{\varepsilon_{ij}}}{\partial a_o}\frac{\partial a_o}{\partial \varepsilon_{lm}}\point
\end{equation}
In case of the Voigt-Taylor approach, $\jumpeps\equiv\bold{0}$ and the derivatives of $\jumpeps$ vanish. For the partial rank-I relaxation, $\jumpeps$ depends on the strain state $\eps$, cf. Appendix~\ref{app:localnr}, which is why the chain rule is applied. The corresponding derivative
\begin{equation}
\label{appeq:deps_da}
\frac{\partial\jump{\varepsilon_{ij}}}{\partial a_o}=\frac{1}{2}\left(\delta_{oi} n^\text{i}_j  +\delta_{oj} n^\text{i}_i\right)\point 
\end{equation}
is obtained by exploiting Equation~\eqref{eq:strjumppr1}. Hence, the last summand in Equation~\eqref{appeq:stressstrain} vanishes, because Equation~\eqref{eq:neccondpr1} holds. This has been confirmed in combination with the local \textsc{Newton-Raphson} scheme~\ref{app:localnr}, which yielded deviations from zero in the order of $\SI{e-20}{}$. Finally,
\begin{equation}
\label{appeq:stressstrainfinal}
\frac{\partial\opsi{ }{el}{ }}{\partial \varepsilon_{lm}} = (1-p) \,\overset{1}{\sigma}_{lm} + p\,\overset{2}{\sigma}_{lm}
\end{equation}
is obtained, which is equal to Equation~\eqref{eq:cauchystress}. 

\section{-- Local \textsc{Newton-Raphson} scheme for the strain jump}
\label{app:localnr}
A local \textsc{Newton-Raphson} scheme is employed in order to determine the strain jump amplitude $\ve{a}$, for which
\begin{equation}
0\overset{!}{=}\left[g(c)\left(\osig{2}{+}-\osig{1}{+}\right)+\osig{2}{-}-\osig{1}{-}\right]\T\cdot\noi
\end{equation}
holds, cf. Equation~\eqref{eq:neccondpr1}. For this purpose, the local residual 
\begin{equation}
^iR^\text{loc}_m=\left[\overset{2}{\sigma}_{lm}-\overset{1}{\sigma}_{lm}\right]n^\text{i}_l\quad\text{with}\quad\overset{i}{\sigma}_{lm}=g(c)\,\overset{i}{\sigma}\mathstrut_{lm}^++\overset{i}{\sigma}\mathstrut_{lm}^-
\end{equation}
of the $i$th local iteration is defined. Index notation is used for better readability of the tensor products. It is noted, that the stresses of the individual phases comprise the degraded and persistent stress contributions. A first order Taylor series expansion of the local residual with respect to a variation of the strain jump amplitude yields
\begin{equation}
^{i+1}R^\text{loc}_m\approx\mathstrut^iR^\text{loc}_m+\left.\frac{\partial R^\text{loc}_m}{\partial a_o}\right|_{i} \mathstrut^{i+1}\Delta a_o \overset{!}{=}0\point
\end{equation}
The incremental increase of the strain jump amplitude 
\begin{equation}
\mathstrut^{i+1}\Delta a_o = -\left(\left.\frac{\partial R^\text{loc}_m}{\partial a_o}\right|_{i} \right)^{-1}\mathstrut^iR^\text{loc}_m\quad\text{with} \quad \mathstrut ^{i+1}a_o=\mathstrut^ia_o+\mathstrut^{i+1}\Delta a_o
\end{equation}
can be obtained by rearrangement. The local tangent reads
\begin{equation}
\label{appeq:apploctang}
\frac{\partial R^\text{loc}_m}{\partial a_o}=n^\text{i}_l\left[(1-p)\,\overset{2}{\mathfrak{C}}_{lmoh}+p\,\overset{1}{\mathfrak{C}}_{lmoh}\right]n^\text{i}_h\quad\text{with} \quad\overset{i}{\mathfrak{C}}_{lmoh}=\frac{\overset{i}{\sigma}_{lm}}{\overset{i}{\varepsilon}_{oh}}=g(c)\frac{\overset{i}{\sigma}\mathstrut_{lm}^+}{\overset{i}{\varepsilon}_{oh}}+\frac{\overset{i}{\sigma}\mathstrut_{lm}^-}{\overset{i}{\varepsilon}_{oh}}\point
\end{equation}
The specific choice of the strain energy split influences the stress tangents $\overset{i}{\mathfrak{C}}_{lmoh}$ of the individual phases. Hence, the scheme can be applied to any split, which fits in the present model. 

For the consistent linearization in Appendix~\ref{app:weaklinear}, the derivative $\partial a_o/\partial \varepsilon_{lm}$ is needed. It is obtained, by implicitly differentiating the converged local residual 
\begin{equation}
^iR^\text{loc}_m=\bold{0}=\left[\overset{2}{\sigma}_{lm}-\overset{1}{\sigma}_{lm}\right]n^\text{i}_l\qquad \left\vert \frac{\partial(\,\,)}{\partial\varepsilon_{op}}  \right.
\end{equation}
with respect to the strain. Together with Equations~\eqref{appeq:eps1eps} and \eqref{appeq:eps2eps} and some rearrangement,
\begin{equation}
\bold{0}=\left(\overset{2}{\mathfrak{C}}_{lmop}-\overset{1}{\mathfrak{C}}_{lmop}\right)n^\text{i}_l+\frac{\partial a_t}{\partial \varepsilon_{op}}\underbrace{n^\text{i}_l\left(p\,\overset{1}{\mathfrak{C}}_{lmts}+(1-p)\,\overset{2}{\mathfrak{C}}_{lmts}\right)n^\text{i}_s}_{A_{mt}}
\end{equation}
is obtained, which finally gives
\begin{equation}
\label{appeq:da_deps}
\frac{\partial a_t}{\partial \varepsilon_{op}}=\left(\overset{1}{\mathfrak{C}}_{lmop}-\overset{2}{\mathfrak{C}}_{lmop}\right)n^\text{i}_l A_{mt}^{-1}\point
\end{equation}
\section{-- Weak form and consistent linearization}
\label{app:weaklinear}
The weak form of the coupled differential equations~\eqref{eq:linmom} and \eqref{eq:pfevo} reads
\begin{equation}
\label{appeq:weak}
\begin{split}
0=&\int\limits_\Omega \sig:\delta \eps \dv-\int\limits_{\partial\Omega_{\ve{t}}} \bar{\ve{t}}\cdot \delta\ve{u}\da\\
&-\int\limits_\Omega \left[ \frac{\mathcal{G}_\text{c}}{2\ellc}-c\left(\frac{\mathcal{G}_\text{c}}{2\ellc}+2\,(1-\eta)\,\psi^\text{el}_{+}\right)-  \eta_\text{f}\frac{c-c_n}{\tau}\right]\delta c - 2\ellc\mathcal{G}_\text{c}\,\nabla c \cdot\nabla\delta c  \dv\coma
\end{split}
\end{equation}
where $\delta\eps=\frac{1}{2}(\nabla\delta\ve{u}+(\nabla\delta\ve{u})\T )=\nabla_\text{s}\delta\ve{u}$, $\delta\ve{u}$ and $\delta c$ denote the variations of the strain, the displacement and the phase-field. The total domain is subsequently divided into $n_\text{el}$ finite elements for which, the primary field variables and corresponding variations are approximated by
\begin{equation}
\lbrace \ve{u},c\rbrace\approx\lbrace N^\alpha \ve{u}^\alpha,N^\alpha c^\alpha \rbrace \quad\text{and}\quad
\lbrace \delta\ve{u},\delta c\rbrace\approx\lbrace N^\alpha \delta\ve{u}^\alpha,N^\alpha \delta c^\alpha \rbrace\coma
\end{equation}
with shape functions $N^\alpha$, where $\alpha$ is the global node number, and the corresponding nodal value $(\bullet)^\alpha$. The weak form~\eqref{appeq:weak} has to hold for arbitrary variations $\delta\ve{u}$ and $\delta c$, which is why the residuals 
\begin{align}
^{n+1,i}\bold{R}^\alpha&=\bigcup_{e=1}^{n_\text{el}}\Bigg[\,  \int\limits_{\Omega^e} \negthickspace\nabla_\text{s}N^\alpha\!\cdot \sig \dv-\int\limits_{\partial\Omega^e_{\ve{t}}} \negthickspace\negthickspace N^\alpha\,\bar{\ve{t}}\da  \Bigg]=\bold{0}\quad\text{and}\\
^{n+1,i}R^\alpha&=\bigcup_{e=1}^{n_\text{el}}\Bigg[\, -\int\limits_{\Omega^e}\negthickspace N^\alpha\!\left[ \frac{\mathcal{G}_\text{c}}{2\ellc}-c\left(\frac{\mathcal{G}_\text{c}}{2\ellc}+2\,(1-\eta)\,\psi^\text{el}_{+}\right)-  \eta_\text{f}\frac{c-c_n}{\tau}\right] - \nabla N^\alpha\, 2\ellc\mathcal{G}_\text{c}\cdot\nabla c  \dv \Bigg]=0
\end{align}
of the $i$th iteration and $(n+1)$th time step are defined. All variables with no time index refer to time step $n+1$. A first order Taylor series expansion of the residuals with respect to a variation of the primary field variables gives
\begin{alignat}{3}
^{n+1,i+1}\bold{R}^\alpha&\approx \mathstrut ^{n+1,i}\bold{R}^\alpha + &\left.\frac{\partial \bold{R}^\alpha}{\partial \ve{u}^\beta}\right|_{n+1,i}& \cdot\mathstrut^{n+1,i+1}\Delta \ve{u}^\beta+ &\left.\frac{\partial \bold{R}^\alpha}{\partial c^\beta}\right|_{n+1,i}& \mathstrut^{n+1,i+1}\Delta c^\beta \overset{!}{=}0 \quad \text{and}\\
^{n+1,i+1}R^\alpha&\approx\mathstrut ^{n+1,i}R^\alpha+ &\left.\frac{\partial R^\alpha}{\partial \ve{u}^\beta}\right|_{n+1,i}&\cdot \mathstrut^{n+1,i+1}\Delta \ve{u}^\beta+ &\left.\frac{\partial R^\alpha}{\partial c^\beta}\right|_{n+1,i} &\mathstrut^{n+1,i+1}\Delta c^\beta \overset{!}{=}0\quad.
\end{alignat}
In order to determine the incremental increase of the displacement and the phase-field for the $(n+1)$th time step, the system of equations
\begin{equation}
\begin{bmatrix}
\bold{K}^{\alpha\beta}_{\bold{u}\bold{u}}&\bold{K}^{\alpha\beta}_{\bold{u}c} \\
\bold{K}^{\alpha\beta}_{c\bold{u}}  & K^{\alpha\beta}_{cc} 
\end{bmatrix}
\begin{bmatrix}
\Delta \bold{u}^\beta\\
\Delta c^\beta
\end{bmatrix}=-\begin{bmatrix}
\bold{R}^\alpha\\
R^\alpha
\end{bmatrix}\quad\text{with}\quad\alpha,\beta=1,2\dots n_\text{nd}
\end{equation}
is solved for the global node number $n_\text{nd}$ in every iteration $i$. It is noted, that the global stiffness matrix is symmetric due to the variational structure of the problem. The four submatrices are given as
\begin{align}
\bold{K}^{\alpha\beta}_{\bold{u}\bold{u}}=\frac{\partial \bold{R}^\alpha}{\partial \ve{u}^\beta}=&\bigcup_{e=1}^{n_\text{el}}\Bigg[\,  \int\limits_{\Omega^e} \negthickspace\nabla_\text{s}N^\alpha\!\cdot \frac{\partial\sig}{\partial\eps} \cdot\nabla_\text{s}N^\beta\dv \Bigg]\quad,\\
\bold{K}^{\alpha\beta}_{\bold{u}c}=\frac{\partial \bold{R}^\alpha}{\partial c^\beta}=&\bigcup_{e=1}^{n_\text{el}}\Bigg[\,  \int\limits_{\Omega^e} \negthickspace\nabla_\text{s}N^\alpha\!\cdot \frac{\partial\sig}{\partial c}  N^\beta\dv \Bigg]=\left(\bold{K}^{\alpha\beta}_{c\bold{u}}\right)\T\quad\text{and}\\
K^{\alpha\beta}_{cc}=\frac{\partial R^\alpha}{\partial c^\beta}=&\bigcup_{e=1}^{n_\text{el}}\Bigg[\, \int\limits_{\Omega^e}\negthickspace N^\alpha\!\left[ \frac{\mathcal{G}_\text{c}}{2\ellc}+2\,(1-\eta)\,\psi^\text{el}_{+}+  \frac{\eta_\text{f}}{\tau}\right]N^\beta + \nabla N^\alpha\, 2\ellc\mathcal{G}_\text{c}\cdot\nabla N^\beta  \dv \Bigg]\coma
\end{align}
with the derivatives
\begin{equation}
\begin{split}
\frac{\partial\sigma_{lm}}{\partial\varepsilon_{ij}}=&\,\,(1-p)\,\overset{1}{\mathfrak{C}}_{lmij}+p\,\overset{2}{\mathfrak{C}}_{lmij}+p\,(1-p)\left\lbrace \left(\overset{2}{\mathfrak{C}}_{lmos}-\overset{1}{\mathfrak{C}}_{lmos}\right)n^\text{i}_s\frac{\partial a_o}{\partial \varepsilon_{ij}} \right.\\ &\left.+\frac{\partial a_o}{\partial \varepsilon_{lm}}n^\text{i}_p\left(\overset{2}{\mathfrak{C}}_{poij}-\overset{1}{\mathfrak{C}}_{poij}\right)+\frac{\partial a_o}{\partial \varepsilon_{lm}}n^\text{i}_p\left((1-p)\,\overset{2}{\mathfrak{C}}_{pors}+p\,\overset{1}{\mathfrak{C}}_{pors}\right)n^\text{i}_s\frac{\partial a_r}{\partial \varepsilon_{ij}}\right\rbrace
\end{split}
\end{equation}
and
\begin{equation}
\frac{\partial\sigma_{lm}}{\partial c}=2c\,(1-\eta)\left[(1-p)\,\overset{1}{\sigma}\mathstrut_{lm}^++p\,\overset{2}{\sigma}\mathstrut_{lm}^++p\,(1-p)\left(\overset{2}{\sigma}\mathstrut_{oq}^+-\overset{1}{\sigma}\mathstrut_{oq}^+\right)n^\text{i}_q\frac{\partial a_o}{\partial \varepsilon_{lm}}\right]\point
\end{equation}
Please refer to Equations~\eqref{appeq:stressstrain}, \eqref{appeq:deps_da}, \eqref{appeq:apploctang} and \eqref{appeq:da_deps} for the evaluation of the two above equations. For better readability of the tensor products, the index notation is used. It is noted, that the evaluation of $\partial^2 a_o/\partial\varepsilon_{lm}\partial\varepsilon_{ij}$ is not necessary, because the stress jump condition is pointwise fulfilled for $p\in(0,1)$.
\section{-- Boundary condition and reference energy for convergence investigation}
\label{app:traction}
A circular disc according to Figure~\ref{fig:circInclGeom} is considered. In a general setting, the inner and outer part have different elastic properties, i.e. $\eo, \nuo$ and $\et, \nut$, respectively. For the plane strain setting, auxiliary material parameters are defined as
\begin{equation}
\eto=\frac{\eo}{1-\nuo^2}\coma \quad\ett=\frac{\et}{1-\nut^2}\coma \quad\nuto=\frac{\nuo}{1-\nuo}\quad \text{and} \quad\nutt=\frac{\nut}{1-\nut}\point
\end{equation}
The radial and tangential strain components are given as
\begin{align}
\varepsilon_{rr}(r)&=\begin{cases}
\frac{1-\nuto}{\eto}K_{11}\quad &\text{for}\quad r\leq r_1\\[2ex]
\frac{1-\nutt}{\ett}K_{12}-\frac{1+\nutt}{\ett} \frac{K_{22}}{r^2}\quad &\text{for}\quad r> r_1
\end{cases}\quad\text{and}\\
\varepsilon_{\varphi\varphi}(r)&=\begin{cases}
\frac{1-\nuto}{\eto}K_{11}\quad &\text{for}\quad r\leq r_1\\[2ex]
\frac{1-\nutt}{\ett}K_{12}+\frac{1+\nutt}{\ett} \frac{K_{22}}{r^2}\quad &\text{for}\quad r> r_1
\end{cases}\quad.
\end{align}
The three integration constants read
\begin{align}
K_{11}&=\left[2r_2u_\text{R}\eto\ett \right]/C\coma\\
K_{12}&=\left[r_2u_\text{R}\ett \left(\eto\left(1+\nutt\right)+\ett\left(1-\nuto\right)\right) \right]/C\quad\text{and}\\
K_{22}&=\left[r_1^2r_2u_\text{R}\ett \left(\ett\left(1-\nuto\right)-\eto\left(1-\nutt\right)\right) \right]/C\coma\\
\text{with}\quad C&=\eto\left(1-\nutt^2\right) \left(r_2^2-r_1^2 \right)+\ett\left(1-\nuto\right) \left(\left(1-\nutt\right)r_2^2+\left(1+\nutt\right)r_1^2 \right)\point
\end{align}
The radial and tangential stress components are obtained from
\begin{align}
\sigma_{rr}(r)&=
\frac{\tilde{E}_i}{1-\tilde{\nu}_i^2}\left(\varepsilon_{rr}(r)+\tilde{\nu}_i\,\varepsilon_{\varphi\varphi}(r)\right)\quad\text{and}\\
\sigma_{\varphi\varphi}(r)&=
\frac{\tilde{E}_i}{1-\tilde{\nu}_i^2}\left(\varepsilon_{\varphi\varphi}(r)+\tilde{\nu}_i\,\varepsilon_{rr}(r)\right)
\end{align}
for $i=1$ if $r\leq r_1$, and  $i=2$ if $r> r_1$. The strain energy within the simulated domain in Figure~\ref{fig:circInclGeom} can be calculated analytically 
\begin{equation}
\label{appeq:energ}
\Psi^\text{el,tot}=K_{11}^2\frac{1-\nuto}{\eto}r_1^2\frac{\pi}{4}+K_{12}^2\frac{1-\nutt}{\ett}\left(a^2-r_1^2\frac{\pi}{4}\right)+K_{22}^2\frac{1+\nutt}{\ett}\left(\frac{\pi}{4r_1^2}-\frac{\pi+2}{8a^2}\right)
\end{equation}
for the square domain of unit thickness, and is used as reference for the convergence investigation of the total energy for the case without a crack. The stress components in Cartesian coordinates read
\begin{align}
\sigma_{xx}(x,y)&=\sigma_{rr}(r)\,\cos^2\negmedspace\varphi+\sigma_{\varphi\varphi}(r)\,\sin^2\negmedspace\varphi\coma\\
\sigma_{yy}(x,y)&=\sigma_{rr}(r)\,\sin^2\negmedspace\varphi+\sigma_{\varphi\varphi}(r)\,\cos^2\negmedspace\varphi\coma\\
\sigma_{xy}(x,y)&=\sigma_{rr}(r)\,\sin\varphi\,\cos\varphi-\sigma_{\varphi\varphi}(r)\,\sin\varphi\,\cos\varphi\coma
\end{align}
with the angle $\varphi=\arctan(y/x)$ and radius $r=\sqrt{x^2+y^2}$. The traction $\bar{\ve{t}}$ along $\partial\Omega_\text{top}$ and $\partial\Omega_\text{right}$ for the convergence investigation with and without a crack is obtained from

\begin{equation}
\begin{pmatrix}
\label{appeq:traction}
\sigma_{xx}(\bar{x},\bar{y}) &\sigma_{xy}(\bar{x},\bar{y})\\
\sigma_{yx}(\bar{x},\bar{y}) &\sigma_{yy}(\bar{x},\bar{y})
\end{pmatrix}\cdot\nob=\bar{\ve{t}}\quad\text{with}\quad \lbrace\bar{x},\bar{y} \rbrace=\lbrace x,y  \,\vert\, x,y \in \partial\Omega_\text{top} \cup \partial\Omega_\text{right} \rbrace\point
\end{equation}

\bibliographystyle{unsrt}      
\bibliography{literatur.bib}   

%
%

\end{document}